\documentclass{article}

\usepackage{graphicx}
\usepackage{dcolumn}
\usepackage{bm}
\usepackage{enumitem}
\usepackage{textcomp}
\usepackage{gensymb}

\usepackage[letterpaper,top=2cm,bottom=2cm,left=3cm,right=3cm,marginparwidth=1.75cm]{geometry}

\usepackage{soul}

\newcommand{\wu}{\small Department of Physics, Wittenberg University, Springfield 45501 OH, USA}
\newcommand{\msu}{\small Facility for Rare Isotope Beams and Department of Physics and Astronomy,
Michigan State University, East Lansing 48824 MI, USA}
\newcommand{\irl}{\small International Research Laboratory on Nuclear
Physics and Nuclear Astrophysics,
Centre National de la Recherche Scientifique - Michigan State University, East Lansing 48824 MI, USA}

\usepackage{xcolor}
\usepackage{amsmath}
\usepackage{physics}
\usepackage{bm}
\usepackage{authblk}

\title{Comments on the publication {\em Discrete symmetries tested at $10^{-4}$ precision
using linear polarization of photons from positronium annihilations} by P.~Moskal et al.\footnote{Submitted to Nature Communications on June 12, 2024}}

\author[1]{E.A.~George}
\author[2]{T.E.~Haugen}
\author[2,3]{O.~Naviliat-Cuncic\footnote{Corresponding author: naviliat@frib.msu.edu}}
\author[1]{P.A.~Voytas}

\affil[1]{\wu}
\affil[2]{\msu}
\affil[3]{\irl}

\date{June 12, 2014}
\begin{document}
\maketitle
The authors of Ref.~\cite{Mos24} (referred to here as ``Moskal et al.'')
claim to have performed the most precise test of
$P$, $T$ and $CP$ invariance in the decay of ortho-Positronium (o-Ps).
Moskal et al. present a {\em ``methodology involving polarization of photons
from these decays''} \cite{Mos24}. In short, given two photons, $i$ and $j$, out
of the three from o-Ps decay, having respectively momentum vectors $\vec{k}_i$ and $\vec{k}_j$, Moskal et al. effectively measured the triple momentum correlation described by the mixed product
\begin{equation}
{\cal O}_2 = (\vec{k}_i \times \vec{k}'_i)\cdot \hat{k}_j/h_i \equiv \hat{n}_i \cdot \hat{k}_j \ ,
\label{eq:O2}
\end{equation}
where $\vec{k}'_i$ is the momentum of photon $i$ after Compton scattering on a
plastic scintillator bar of the JPET detector,
$\hat{n}_i = (\vec{k}_i \times \vec{k}'_i)/h_i$ is the unit vector perpendicular to
the Compton scattering plane, with $h_i = |\vec{k}_i \times \vec{k}'_i|$ the norm
of the vector product.

In this note: 1) we demonstrate, assuming standard properties for
Compton scattering, that the average value of ${\cal O}_2$ must necessarily be zero, independently
of any physics occurring in o-Ps decay; 2) we point out that there is no formal justification
to equate the normal vector to the Compton scattering plane with the incident photon
polarization, as done by Moskal et al.; 3) we observe the absence of characterization of the device as a Compton polarimeter, which is paramount in photon polarimetry;
4) we review previous measurements of the polarization of photons from o-Ps decay, properly
implementing the Compton polarimetry technique, and make the connection
with tests of discrete symmetries; and
5) we stress that the proposed correlation cannot be generated solely by the physics of
o-Ps decay, including possible violations of discrete symmetries.

\bigskip
1) Moskal et al. report the measurement of the average value of the correlation given
in Eq.(\ref{eq:O2}) and claim this to be a test of $P$, $T$ and $CP$ violation in o-Ps decay.
The average value is
\begin{equation}
\langle {\cal O}_2 \rangle= \langle \hat{n}_i\cdot \hat{k}_j \rangle = \langle cos(\omega_{ij}) \rangle \ ,
\label{eq:avg_O2}
\end{equation}
and the measured distribution of values of ${\cal O}_2$ is shown in Fig.2 of Ref.~\cite{Mos24}.
Now, the Klein-Nishina cross section for Compton scattering is given by \cite{Fag59}
\begin{equation}
 \frac{d\sigma}{d\Omega} = \frac{1}{2}r_e^2 s_e^2 \left[s_e + s_e^{-1} - 2\sin^2\theta \cos^2\phi \right] \ ,
\label{eq:KN_xSec}
\end{equation}
where $r_e = e^2/m_e c^2$ is the classical electron radius,
$s_e = [1 + (E_\gamma/m_ec^2)(1-\cos\theta)]^{-1}$ is the ratio between the energy of
the scattered photon, relative to the energy $E_\gamma$ of the incident photon,
$\theta$ is the angle between $\vec{k}'_i$ and $\vec{k}_i$ (i.e. the Compton scattering angle), and $\phi$ is the angle of
the photon polarization plane relative to the Compton scattering plane.
It is clear that, independently of the photon polarization, the scattering probabilities
at angles $\theta$ and $-\theta$ are identical. This means that one expects to observe equal
numbers of events with $\vec{k}_i \times \vec{k}'_i$ pointing in some direction and events
with this product pointing in the
opposite direction. That is, for a given momentum $\vec{k}_j$, one expects to observe equal
numbers of events producing a positive value of the projection
\begin{equation}
\hat{n}_i \cdot \hat{k}_j = \cos(\omega_{ij})\ ,
\label{eq:cos_omega}
\end{equation}
and a negative value of this projection, due to the change of sign of $\theta$
and hence of the direction of $\hat{n}_i$.
In conclusion, based on the
symmetry properties of the Klein-Nishina cross section, the average value
$\langle {\cal O}_2 \rangle$ must be zero by construction, regardless of any physics in
o-Ps decay.

\bigskip
2) Moskal et al. assert in the title, the abstract, several times in the text
and in the discussion, having used the linear polarization of photons to
perform the measurement. Moskal et al.
introduce in Eq.(2) of Ref.~\cite{Mos24}, the {\em ``polarization related quantities''}
$\bm{\epsilon}_i$ which, later in the text, are simply referred to as the photon polarization.
The vector $\bm{\epsilon}_i$ introduced by Moskal et al. is in fact the vector
$\hat{n}_i$ defined in Eq.~(\ref{eq:O2}) above,
which is the unit vector normal to the Compton scattering plane.
In other words, Moskal et al. equate the polarization of the
incident photon to the normal to the Compton scattering plane.
However, there is no formal justification based on QED to make this association.
Furthermore, we stress that a single Compton scattering event does not enable
the determination of the direction of the polarization of a photon.
This observation is at variance with the statement made in the abstract of
Ref.~\cite{Mos24}, that the photon polarization was
determined {\em ``on an event-by-event basis''}.
In conclusion, the unit vector perpendicular to the Compton scattering plane, which is
called ``photon polarization'' in Ref.~\cite{Mos24}, is not the photon polarization.

\bigskip
3) Compton polarimetry is a well established technique to determine the photon linear
polarization \cite{Fag59}.
It relies on the dependence of the Klein-Nishina cross section, Eq.(\ref{eq:KN_xSec}),
as a function of $\phi$. The $\cos^2\phi$ dependence indicates that there is no
meaningful distinction to
make between something like a ``positive'' versus a ``negative'' photon polarization
direction, since the cross section is invariant under the transformation
$\phi \rightarrow \phi + \pi$. The conception
of Moskal et al., in which the photon polarization has an {\bf oriented} direction,
is inconsistent with this invariance property.
Furthermore, measurements of the photon polarization strictly require
the determination of the analyzing power of the polarimeter which defines the sensitivity
of the device, including variations with angular and energy ranges.
A recent illustration
of the Compton polarimetry technique, using a $\gamma$-tracking multi-detector
array can be found in Ref.~\cite{Mor22}.
We observe that the text in Ref.~\cite{Mos24} does not contain any consideration
about the sensitivity of the setup for the measurement of the photon polarization.
This is possibly connected to the fact that
Moskal et al. call ``photon polarization'' the normal vector to the Compton
scattering plane. This normal vector can indeed be determined from the topology
of a single scattering event but the polarization of the incident photon cannot.
In conclusion, despite the insistence made in Ref.~\cite{Mos24} on the photon polarization,
Moskal et al. did not measure any photon polarization.

\bigskip
4) Measurements of the polarization of photons from o-Ps decay \cite{Lei53,Ye88},
properly using the Compton polarimetry technique, provide
tests of QED and in particular the predictions of the Klein-Nishina formula.
It is not excluded that measurements of correlations involving the photon
polarization could probe form factors in o-Ps decay which violate discrete symmetries.
To our
knowledge this has so far not been demonstrated and would require the extension of
the formalism in Ref.~\cite{Ber88} to include the photon polarizations.

\bigskip
5) The three momenta in Eq.(\ref{eq:O2}) do not originate from the same process:
$\vec{k}_i$ and $\vec{k}_j$ are involved in o-Ps decay whereas $\vec{k}_i$ and $\vec{k}'_i$
are involved in Compton scattering.
The mere presence of vector $\vec{k}'_i$ clearly shows
that the dynamics of o-Ps decay alone will never produce such a correlation.
QED calculations in o-Ps decay, allowing for possible
discrete symmetries violations \cite{Ber88}, can produce symmetry-violating
form factors which are manifested
by correlations involving kinematic quantities of the initial and final states,
like the o-Ps spin, $\bm{s}$,
the photon momenta, $\bm{k}_i$ or the photon polarizations $\bm{e}_i$. The question
about how to measure a correlation in order to access the form factor is
independent and can envision different experimental techniques. 
Since the triple momentum correlation in Eq.(\ref{eq:O2}) cannot be generated by the
dynamics of o-Ps decay, including possible violations of discrete symmetries,
there is nothing it can tell us about any form factor from the decay.

\bigskip
To summarize, it appears that point 1) alone invalidates the claims made in
Ref.~\cite{Mos24} concerning the tests of discrete symmetries. Points 2), 3) and 4)
indicate the misconception by Moskal et al. concerning the photon polarization and hence
how to measure it by Compton polarimetry and what can possibly be learned from a
proper measurement in o-Ps decay. Finally point 5) emphasizes the general properties
to probe the presence of symmetry-violating terms in decay processes and concludes
that the proposed correlation term cannot be generated by the dynamics of o-Ps alone.

\bibliographystyle{CP_comment}
\bibliography{CP_comment} 
\end{document}